\documentclass{article}
\usepackage{amsthm}
\usepackage{spconf,amsmath,epsfig}
\usepackage{amssymb}
\usepackage{array}

\theoremstyle{plain}

\theoremstyle{definition}

\usepackage{subfig}
\usepackage{graphicx}
\usepackage{capt-of}

\usepackage{blindtext}

\DeclareCaptionLabelFormat{andtable}{#1~#2  \&  \tablename~\thetable}

\usepackage{algorithm}
\usepackage{algorithmic}
\usepackage{setspace}
\usepackage{epstopdf}
\usepackage[export]{adjustbox}
\usepackage{subfig}
\usepackage{graphicx}
\usepackage{stfloats}
\usepackage{filecontents,lipsum}
\usepackage[noadjust]{cite}
\def\bt{{\mathbf t}}
\def\bz{{\mathbf z}}
\def\bd{{\mathbf d}}
\def\balpha{{\boldsymbol\alpha}}

\def\bz{{\mathbf z}}

\def\cT{{\mathcal T}}

\def\cN{{\mathcal N}}

\def\cE{{\mathcal E}}
\title{Bayesian Inference of Diffusion Networks with Unknown Infection Times}
\name{Shohreh Shaghaghian %,~\IEEEmembership{Member,~IEEE,}
        and~Mark Coates}
\address{Department of Electrical and Computer Engineering, McGill University}
\begin{document}
%\ninept
%
\maketitle
\begin{abstract}
  The analysis of diffusion processes in real-world propagation
  scenarios often involves estimating variables that are not directly
  observed. These hidden variables include parental relationships, the
  strengths of connections between nodes, and the moments of time that
  infection occurs. In this paper, we propose a framework in which all
  three sets of parameters are assumed to be hidden and we develop a
  Bayesian approach to infer them. After justifying the model
  assumptions, we evaluate the performance efficiency of our proposed
  approach through numerical simulations on synthetic datasets and
  real-world diffusion processes.
\end{abstract}
\begin{keywords}
Information Diffusion, MCMC Inference, Gibbs Sampling
\end{keywords}
\section{Introduction}
\label{sec:intro}

Propagation of information has been widely studied in various domains. Despite different applications \cite{guille2013information}, all of the diffusion models have three main components in common: {\em Nodes}, i.e., the set of separate agents; {\em Infection}, i.e., the change in the state of a node that can be
transferred from one node to the other; and {\em Causality}, i.e., the
underlying structure based on which the infection is transferred between nodes. The term {\em cascade} is usually used to refer to the temporal traces left by a diffusion process. One of the main goals of studying the diffusion process is to infer the causality component using observations related to the cascades. While the majority of studies (e.g.~\cite{ gomez2011uncovering, gomez2013structure, embar2014bayesian, farajtabar2015coevolve}) assume that the cascades are perfectly observed, some studies (\cite{sefer2015convex, sadikov2011correcting, amin2014learning,  lokhov2015efficient,farajtabar2015back}) investigate scenarios in which the cascade trace is not directly observable or is at least partially missing. Two important examples of such scenarios are the outbreak of a contagious disease (with nodes being geographical regions) and the impact of external events on the stock returns of different assets. These studies either assume that some portion of the cascade data is observable and try to infer the causality structure from this portion (e.g.~\cite{amin2014learning, lokhov2015efficient, farajtabar2015back}) or infer the structure using some other observable property of the cascade without inferring the cascade trace itself (e.g.~\cite{sefer2015convex}). In this paper, we assume that the cascade or more precisely the infection times cannot be directly observed. Instead, we observe time series with statistics that change as a function of the true infection times.
As opposed to the previous works, we intend to not only infer the causality structure but also  estimate the unobserved infection times. 

Another related body of literature addresses time series segmentation and investigates the techniques of detecting single (\cite{eckley2011analysis}) and multiple (\cite{killick2012optimal})
changepoints in univariate (\cite{fearnhead2005exact,fearnhead2006exact}) and multivariate (\cite{xuan2007modeling, matteson2014nonparametric}) time series. Some of these methods involve an underlying graphical model that captures the correlation structure between the time series, but there is no notion of an infection network. In this paper, we develop a framework in which the infection times, parental relationships and link strengths can be estimated by simultaneously modeling the network structure and performing time series segmentation.

The paper is organized as follows. In the rest of this section, we
briefly review related work. In Section \ref{sec:system_model}, we describe
our system model and formulate the diffusion problem. We present our
proposed inference approach and discuss its modeling assumptions. We
evaluate the performance of our suggested approach using both
synthetic and real world datasets and present the simulation results
in Section \ref{sec:res}. The concluding remarks are made in Section
\ref{sec:conc}.

{\em \bf Related Work:} Most of the earlier work exploring techniques for inferring the structure of an infection or diffusion network assumed that cascades were perfectly observed.
%\cite{yang2010modeling} assumes that the
%diffusion of information is achieved by the influence %of individual
%nodes and learns the influence matrix by modeling the %number of newly
%infected nodes as a function of the influence of %previously infected
%nodes. \cite{Wang2012Sparse} uses the same linear %modeling but takes
%into account the nodes' centralities and introduces %sparsity in the
%estimated influence matrix through regularization %penalties. 
\cite{gomez2011uncovering} proposes a
generative probabilistic model of diffusion that aims to realistically
describe how infections occur over time in a static network. The
infection network and transmission rates are inferred by maximizing the likelihood of an observed set of
infection times. \cite{gomez2013structure}
investigates the diffusion problem in an unobserved dynamic network
for the case when the dynamic process spreading over the edges of the
network is observed. Stochastic convex optimization is employed to infer the dynamic network. \cite{embar2014bayesian} proposes a Bayesian
framework to estimate the posterior distribution of connection
strengths and information diffusion paths given the observed infection
times. \cite{farajtabar2015coevolve} studies the creation of new links in the diffusion network and proposes a joint continuous-time model of information diffusion and network evolution.

In contrast to the work described above, some studies focus on the problem where cascades are not perfectly observed. In~\cite{sefer2015convex}, it is assumed that the partially observed probabilistic information about the state of each node is provided, but the exact state transition times are not observed. These transition times are related to the observed trace via the noise dynamics function. The underlying network is inferred by minimizing the expected loss over all realizations of the observable trace. \cite{amin2014learning} studies the theoretical learnability of tree-like graphs in a setting where only the initial and final states are observed. The goal in~\cite{lokhov2015efficient} is to reconstruct the so called node couplings using dynamic message-passing equations when the cascade observations are only partially available. \cite{farajtabar2015back} develops a two-stage framework to identify the infection source when the node infections are only partially observed and the diffusion trace is incomplete. This paper is categorized in this second group of studies by proposing an approach to simultaneously infer the structure and casacade trace of a diffusion process.
\section{System Model and Inference Procedure} \label{sec:system_model} We consider a set of
$N$ nodes $\cN=\{1,\dots,N\}$ and assume that node $s \in \cN$ is the
source of a contagion $C$ which is transmissible to other nodes of the
network. When $C$ is transferred from node $j$ to node $i$ ($i,j \in
\cN$), we say node $i$ is infected by node $j$. In this case, we refer
to node $j$ as the parent of node $i$, and denote it by $z_i$. We
model this infection process by a directed, weighted graph
$G=(\cN,\cE,\balpha_{N \times N})$ where $\cE$ is the set of weighted
edges, and $\alpha$ is a $N\times N$ link strength matrix. Component
$\alpha_{ij}$ of this matrix denotes the strength of the link between
two nodes $i$ and $j$. A directed edge $j\rightarrow i$ exists if and
only if $z_i=j$. The set of potential parents for node $i$ is denoted
by $\pi_i$ (i.e. $z_i \in \pi_i$).  The definitions of parents and
candidate parents simply implies that $\forall j \in \pi_i : t_j <
t_i$ and $\forall j \notin \pi_i : \alpha_{ij}=0$.

As mentioned in Section \ref{sec:intro}, we focus on the scenarios
where none of the main infection parameters (link strengths, parents,
and infection times) are directly observed. We assume that the only
observation we get from an arbitrary node $i \in \cN$ is a discrete
time signal of length $T$ denoted by $\bd_i=\{d_i^n\}_{n=1:T}$. We
denote the set of all observed time signals by
$\bd=(\bd_1,\dots,\bd_N)$. The goal is to infer the infection parameters
$(\bz,\bt,\balpha)$ that best explain the received signal vector $\bd$
where $\bz=(z_1,\dots,z_N)$ and $\bt=(t_1,\dots,t_N)$. More precisely,
we aim to find the most probable set of parameters
$(\bz^*,\bt^*,\balpha^*)$ conditioned on the received signals $\bd$,
i.e.
\begin{equation}\label{mainopt}
(\bz^*,\bt^*,\balpha^*)= \underset{(\bz,\bt,\balpha)}{\arg \max} \quad f(\bz,\bt,\balpha|\bd)
\end{equation}

In order to solve \eqref{mainopt}, we need to first derive the joint conditional distribution $f(\bz,\bt,\balpha|\bd)$. Using Bayes' rule we have,
\begin{equation}\label{Bayes}
f(\bz,\bt,\balpha|\bd)=\frac{f(\bd|\bt,\bz,\balpha)f(\bt|\bz,\balpha)f(\bz|\balpha)f(\balpha)}{f(\bd)}
\end{equation}
We consider proper prior distributions for components of equation \eqref{Bayes}. As justified in \cite{embar2014bayesian}, we assume that link strengths $\alpha_{ij}$ are independent and model their probability distribution by a Gamma distribution with parameters $a_{ij}$ and $b_{ij}$ i.e. $\alpha_{ij} \sim \Gamma(a_{ij},b_{ij})$. Therefore, 
\begin{equation}\label{prior_alpha}
f(\balpha)=\prod_{i\in \cN , j \in \pi_i} f(\alpha_{ij})= \prod_{i\in \cN , j \in \pi_i} \frac{x^{a_{ij}-1}e^{-\frac{x}{b_{ij}}}}{\Gamma(a_{ij})b_{ij}^{a_{ij}}}
\end{equation}
We also assume that conditioned on the link strengths, the nodes' parents are independent and follow multinomial distributions i.e. 
\begin{equation}\label{prior_z}
f(\bz|\balpha)=\prod_{i \in \cN} f(z_i|\alpha_{ij_{j \in \pi_i}})=\prod_{i \in \cN} \frac{\alpha_{iz_i}}{\sum_{j\in\pi_i}\alpha_{ij}}
\end{equation}
The next step is to consider a proper prior conditional distribution for infection times. As proposed in \cite{gomez2012inferring}, we assume $t_i$ follows an exponential distribution with parameter $\alpha_{iz_i}$. Without loss of generality, we can assume that $t_1 \geq t_2 \geq \dots\geq t_N$. Therefore, 
\begin{equation}\label{prior_tc}
\begin{aligned}
f(\bt|\bz,\balpha)&=\prod_{i \in \cN}f(t_i|\bz,\balpha,t_{i+1:N})=\prod_{i \in \cN} \alpha_{iz_i} e^{-\alpha_{iz_i}(t_i-t_{z_i})}
\end{aligned}
\end{equation}
Finally, we assume that node $i$'s observed data, $\bd_i$, is independent of the observations from other nodes and that it follows two different distributions before and after being infected at $t_i$. Hence,
\begin{equation}\label{prior_d1}
f(\bd|\bz,\bt,\balpha)=\prod_{i \in \cN}f(\bd_i|t_i)
\end{equation}

With the proposed distributions in
\eqref{prior_alpha}-\eqref{prior_d1}, we can calculate the probability
of any arbitrary set $(\bz_0,\bt_0,\balpha_0)$ up to a constant
$\frac{1}{f(\bd)}$ using \eqref{Bayes}. Since the optimization problem
of \eqref{mainopt} cannot be easily solved, we use MCMC methods to
sample from a probability distribution $f(\bz,\bt,\balpha|\bd)$. We
use Gibbs sampling to generate samples of this posterior
distribution. In other words, we use full conditional distributions
for each of the infection parameters $t_i$, $z_i$, $\alpha_{ij}$ ($i,j
\in \cN$) to generate samples. We denote the parents and infection
times of all the nodes in the network except node $i$ respectively by
$\bz_{\overline{i}}$, $\bt_{\overline{i}}$. Also, the link strength of
all the possible links except the link between nodes $i$ and $j$ is
denoted by $\balpha_{\overline{ij}}$. Using Bayes' rule, the full
conditional probablities for Gibbs sampling are:\\
 (a) For parent of an node $i$,
\begin{equation}\label{lem1_1}
f(z_i|\bd,\bz_{\overline{i}},\bt,\balpha)
\propto f(t_i|z_i,\alpha_{iz_i},t_{z_i})f(z_i|\alpha_{ij_{j \in \pi_i}})
\end{equation}
(b) For infection time of an node $i$,
\begin{equation}\label{lem1_2}
f(t_i|\bd,\bz,\bt_{\overline{i}},\balpha)
 \propto f(d_i|t_i)f(t_i|z_i,\alpha_{iz_i},t_{z_i})\prod_{k\in C_i}f(t_k|\alpha_{ki},t_i)
\end{equation}
(c) For link strength between nodes $i$ and $j \in \pi_i$,
\begin{equation}\label{lem1_3}
f(\alpha_{ij_{ j \in \pi_i}}|\bd,\bz,\bt,\balpha_{\overline{ij}})\propto f(t_i|z_i,\alpha_{iz_i},t_{z_i})f(z_i|\alpha_{ij_{j \in \pi_i}})f(\alpha_{ij})
\end{equation}
We evaluate the proficiency of the proposed inference approach in Section \ref{sec:res}.

\section{Simulation Results}\label{sec:res} 

\subsection{Synthetic Data}

We generate a dataset based on the model
\eqref{prior_alpha}-\eqref{prior_d1}. We first randomly choose
$\pi_i$s (for all $i\in\cN$) and an underlying directed tree $\cT$
with adjacency matrix $\mathbf{A}=[A_{ij}]$, where $A_{ij}=1$ if and only if
there is a directed edge from $i$ to $j$. The link strength value
${\alpha_{ij}}$ (${j \in \pi_i}$) is generated using the gamma
distribution $\Gamma (a_1,b_1)$ if $A_{i,j}=1$ and $\Gamma (a_2,b_2)$
if $A_{i,j}=0$. We refer to these $\alpha$ values as \textit{true
  alphas} and denote them by $\balpha^R=[\alpha_{ij}^R]_{N \times N}$.
Then, we choose the parent of node $i$ i.e. $z_i$ from all the nodes
$j \in \pi_i$ based on a random sampling with weights
$\alpha_{ij}$. These parents are called \textit{true parents} and are
denoted by $\bz^R=(z_1^R,\dots,z_N^R)$. Knowing the values of $z_i$
and $\alpha_{iz_i}$, we then generate the \textit{true
  infection times} $\bt^R=(t_1^R,\dots,t_N^R)$ based on the exponential
distributions described in \eqref{prior_tc}. Finally, we generate the data $d_i$ based on two different Gaussian distributions with
parameters $(\mu_{1i},\sigma_{1i})$ and $(\mu_{2i},\sigma_{2i})$ for
all nodes $i \in \cN$, i.e.
\begin{equation}\label{prior_d2}
\begin{aligned}
f(d_i|t_i)=\frac{e^{-[\frac{\sum_{n=1}^{t_i-1} (d_i^n-\mu_{1i})^2}{2\sigma_{1i}^2}+\frac{\sum_{n=t_i}^{T} (d_i^n-\mu_{2i})^2}{2\sigma_{2i}^2}]}}{\sqrt{2\pi}^T \sigma_{1i}^{t_i}\sigma_{2i}^{T-t_i}} 
\end{aligned}
\end{equation}

We generate $M$ samples using full conditional distributions of
equations \eqref{lem1_1}-\eqref{lem1_3} to infer the network
parameters $(\bz^R,\bt^R,\balpha^R)$. We denote the set of all
generated samples by $\mathcal{M}$ and refer to the $m$th sample as
$S^m$. The parent vector, infection time vector, and strength matrix
of the $m$th sample are respectively denoted by $S_{\bz}^m$,
$S_{\bt}^m$, and $S_{\balpha}^m$. Denoting the most observed
parent-infection time pair (i.e. the pair that has been repeated the
most among the $M$ generated samples) by $(\hat{\bz},\hat{\bt})$, we
estimate the components of the link strength
$\hat{\balpha}=[\hat{\alpha}_{ij}]_{N\times N}$ by
$\hat{\alpha}_{ij}=\frac{1}{|\mathcal{S}|} \sum_{k \in \mathcal{S}}
[S_{\balpha}^k]_{ij}$ where $\mathcal{S}=\{m\in
\mathcal{M}|S_{\bz}^m=\hat{\bz},S_{\bt}^m=\hat{\bt}\}$. 
In order to
evaluate the performance of our proposed inference approach, two main
questions should be answered: (1) Does the network structure improve
detection of infection times? (2) How much accuracy is lost in terms
of detecting the parents and estimating link strengths when time
series are observed instead of the actual infection times?

\begin{figure}
\centering
\begin{minipage}{.15\textwidth}
\centering
\vspace*{0.3cm}
\small
    \begin{tabular}{|c|c|c|} 
    	\hline
    \begin{tabular}{@{}c@{}}Sce- \\nario\end{tabular} 
     & $\mu_2$ & $b_2$ \\
	\hline
	{\em A} & $100$ & $0.9$ \\
	\hline
	{\em B} & $100$ & $0.6$ \\
	\hline
	{\em C} & $11$ & $0.9$  \\
	\hline
	{\em D} & $11$ & $0.6$ \\
	\hline
	\end{tabular}
	\captionof{table}{\\Test Scenarios}
	\label{table:scenarios}
\end{minipage}
\hspace{0.4 cm}
\begin{minipage}{.29\textwidth}
\centering
\includegraphics[width=\textwidth]{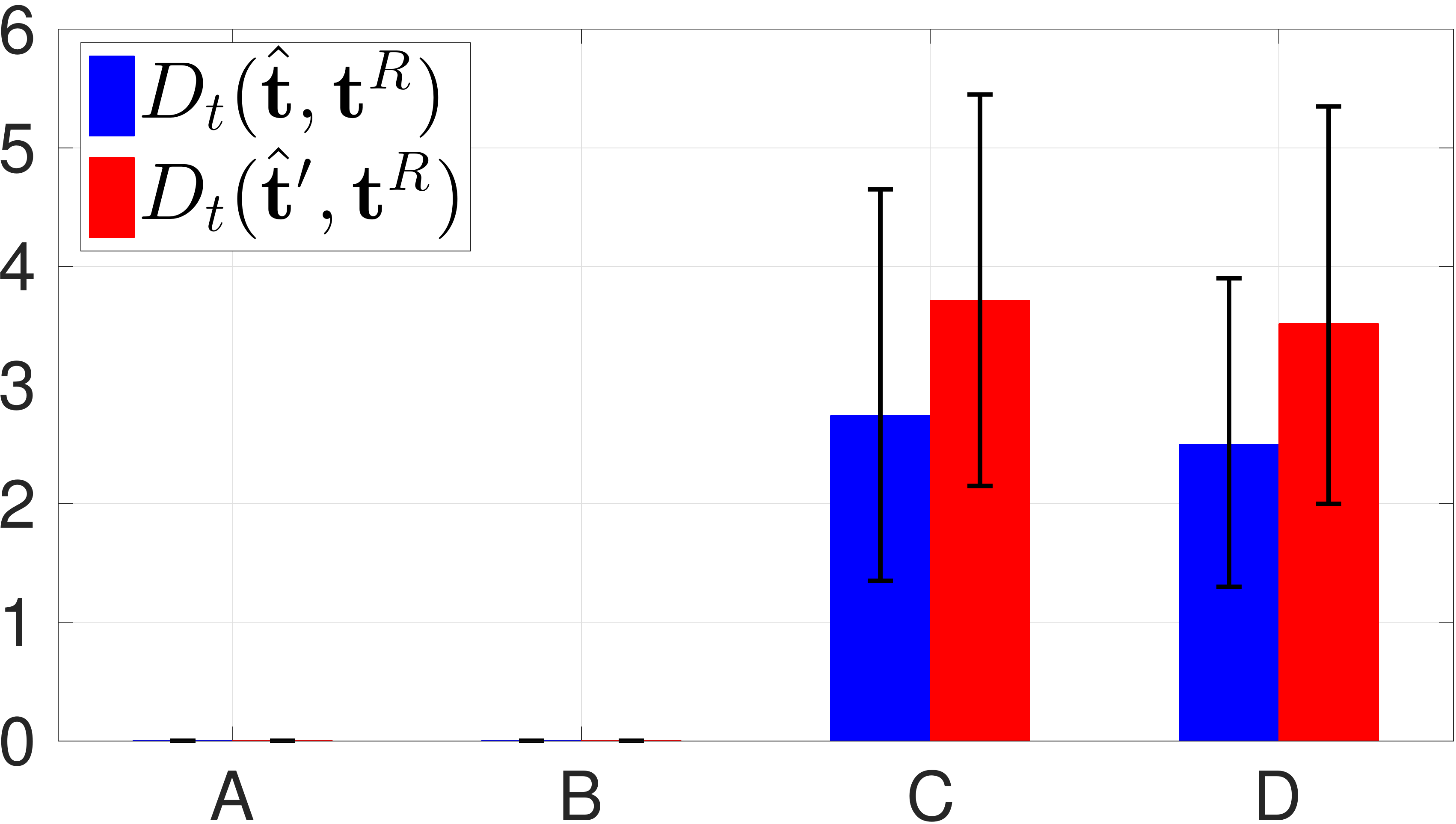}
\caption{Infection Time Deviations}
\label{fig:fig1}
\end{minipage}
\end{figure}

The first question can be answered by comparing the accuracy of
infection time estimates for two cases. In the first case, we detect the
infection time of each node independently (i.e. $\hat{t_i} '=\arg
\max_{t_i} f(t_i|d_i)$), while in the second case we exploit the
network structure to find the infection times as explained in Section
\ref{sec:system_model}. We denote the vector of all $\hat{t_i}'$s by
$\hat{\bt}'$ and define the infection time deviation function
$D_t(\bt_x^1,\bt_x^2)$ as the average number of samples that are
different in the arbitrary infection time vectors
$\bt_x^1=({t_x^1}_1,\dots.{t_x^1}_N)$ and
$\bt_x^2=({t_x^2}_1,\dots.{t_x^2}_N)$ i.e.
%\begin{equation}
$\forall \bt_x^1 , \bt_x^2 \in \mathcal{R}^{1 \times N}  : \quad D_t(\bt_x^1,\bt_x^2)  \triangleq \frac{1}{N} \sum_{i=1}^N |{t_x}^1_i-{t_x}^2_i|$.
%\end{equation}
Figure \ref{fig:fig1} shows the average and $95\% $ confidence intervals of deviation values for
$\bt_x^1=\hat{\bt},\hat{\bt}'$ and $\bt_x^2={\bt}^R$ in $100$
networks of $N=20$ nodes using four extreme sets of parameters
described in Table \ref{table:scenarios}. In all these scenarios, $\mu_{i1}=10$ , $\mu_{i2}=\mu_2$, $\sigma_{i1}=\sigma_{i2}=1$ for all $i\in \mathcal{N}$ and $a_1=9$, $b_1=0.5$ , $a_2=10$. $M=10^5$ samples are
generated and the first $10^3$ generated samples are discarded. As we see in Figure \ref{fig:fig1}, in scenarios {\em A}
and {\em B}, infection times can be detected with high likelihood thus
both performance metrics are zero. However, in scenarios {\em
  C} and {\em D}, we see that exploitation of the network structure
results in smaller deviation from the true values. The infection time
estimates are in average more accurate. 
%\blindtext

We now compare our proposed framework with a more idealized situation
in which the $\bt_i$ values are known. We denote the parents and link
strengths estimated with knowledge of the infection times by $\hat{\bz}'$ and
$\hat{\balpha}'$ and define deviation functions
$D_z(\bz_x^1,\bz_x^2)$ and $D_{\alpha}(\balpha_x^1,\balpha_x^2)$. The
parent deviation function $D_z(\bz_x^1,\bz_x^2)$ is defined as the
number of nodes whose parents are different in
$\bz_x^1=({z_x^1}_1,\dots,{z_x^1}_N)$ and
$\bz_x^2=({z_x^2}_1,\dots,{z_x^2}_N)$ i.e.
%\begin{equation}
$\forall \bz_x^1 , \bz_x^2 \in \mathcal{R}^{1 \times N}  :  \quad D_z(\bz_x^1,\bz_x^2) \triangleq \sum_{i=1}^N I({z_x^1}_i-{z_x^2}_i)$
%\end{equation}
, where $I(x)=1$ if $x\neq0$ and $I(x)=0$ otherwise. Finally, for the deviation of link strengths we have,
%\begin{equation}
$\forall \balpha_x^1 , \balpha_x^2 \in \mathcal{R}^{N \times N}:  D_{\alpha}(\balpha_x^1,\balpha_x^2) \triangleq \frac{1}{N}\sum_{\hat{\alpha}_{ij}>0} |{\alpha_x^1}_{ij}-{\alpha_x^2}_{ij}|$.
%\end{equation}
\begin{figure}[!hbt]
\centering
\includegraphics[width=0.45 \textwidth , height=0.2 \textwidth]{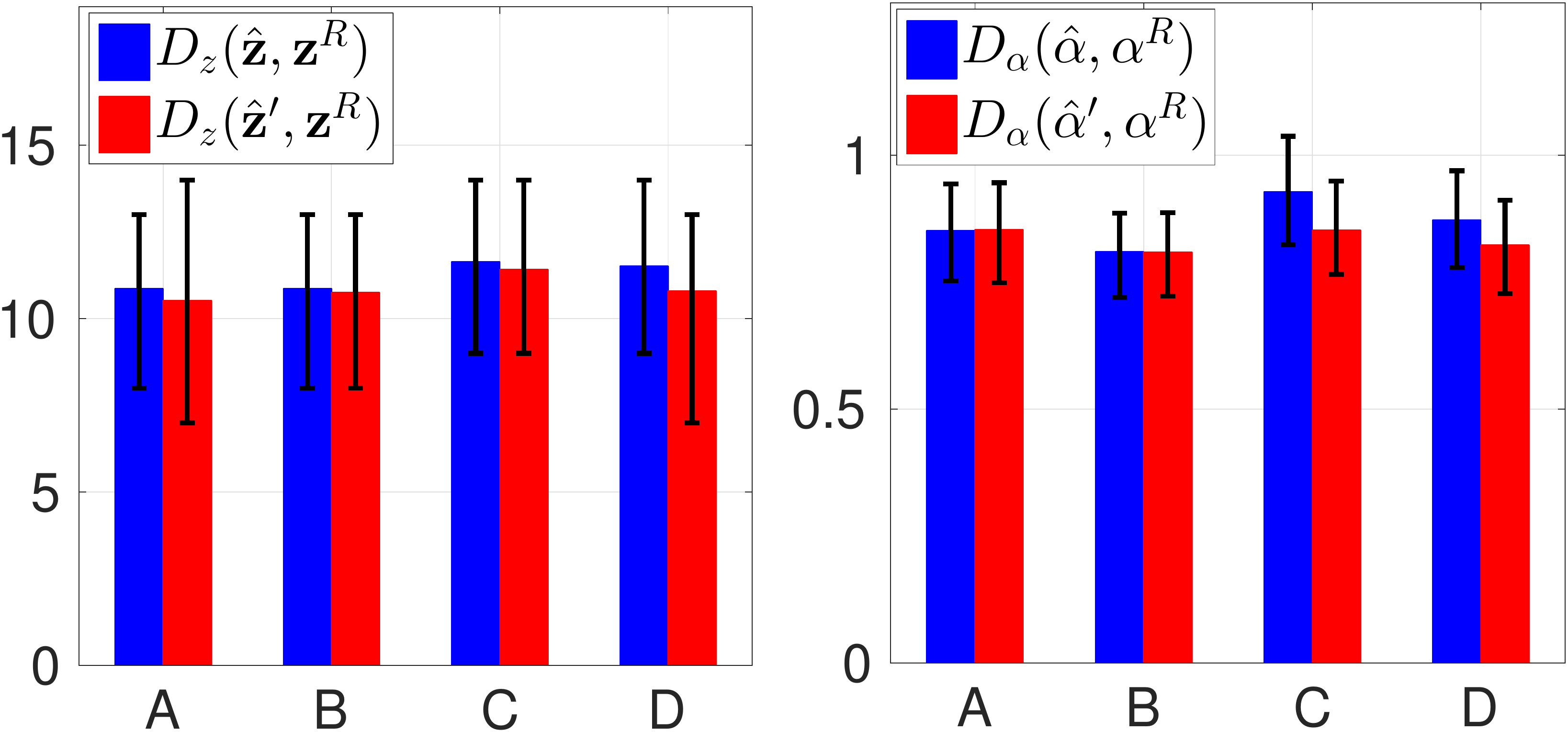}
\caption{Deviations in Detection of $\bz$ and $\balpha$}
\label{fig:fig2}
\end{figure}

Figure \ref{fig:fig2} shows the values of the defined performance
metrics for $\bz_x^1=\hat{\bz},\hat{\bz}'$ and
$\bz_x^2=\bz^R$. We see that in scenarios {\em C} and {\em D}
(where the noise is greater and infection times are more difficult to estimate), not knowing the exact infection times results in larger deviations in estimating the network parameters. Overall, however, the deterioration
in estimation accuracy is not dramatic.

\subsection{Real Data}
\begin{figure}[!htb]
\includegraphics[width=0.47 \textwidth]{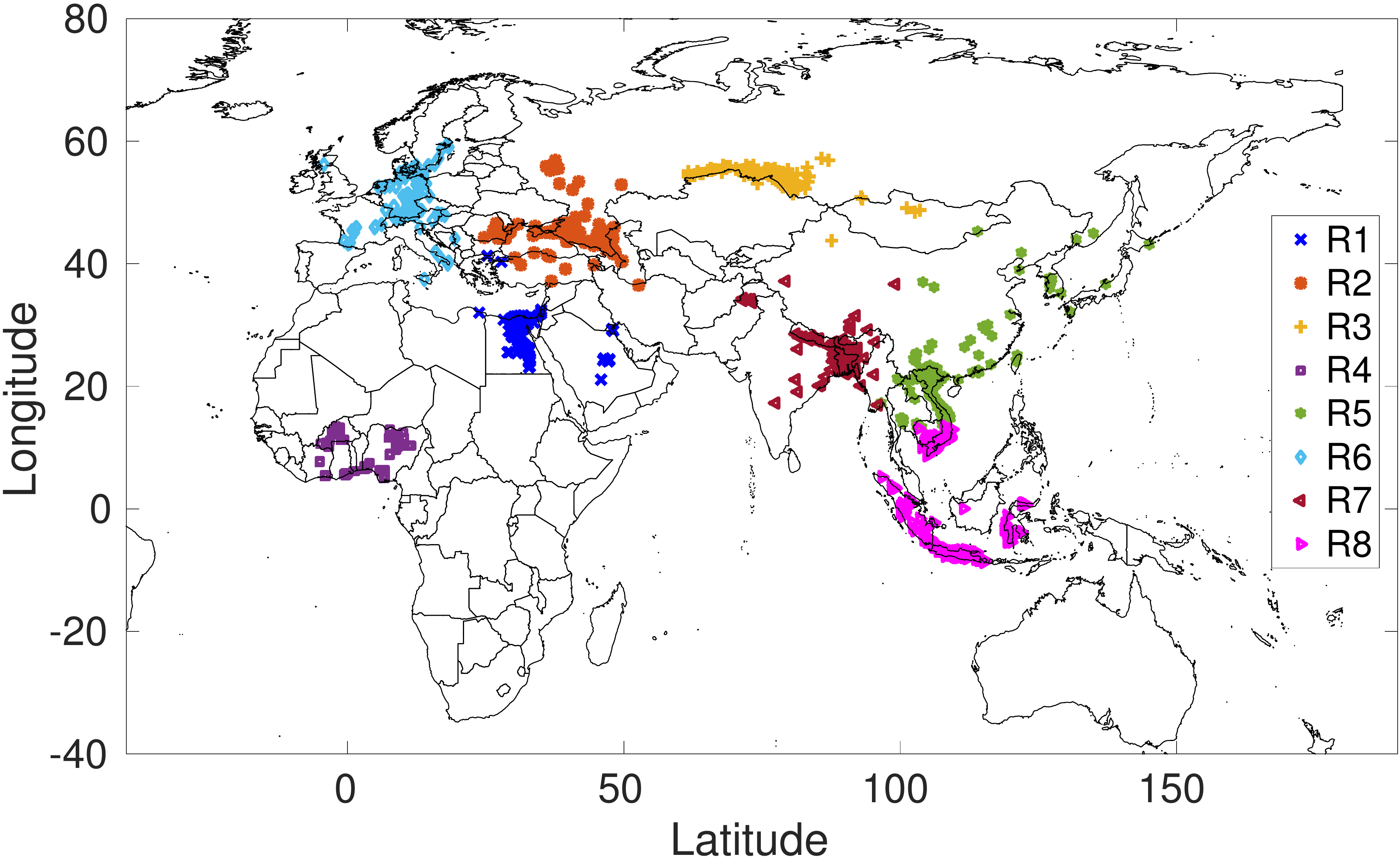}
\caption{H5N1 HPAI outbreak in 2004-2016}
\label{fig:fig3}
\end{figure}

\begin{figure}[!htb]
\centering
\includegraphics[width=0.49 \textwidth]{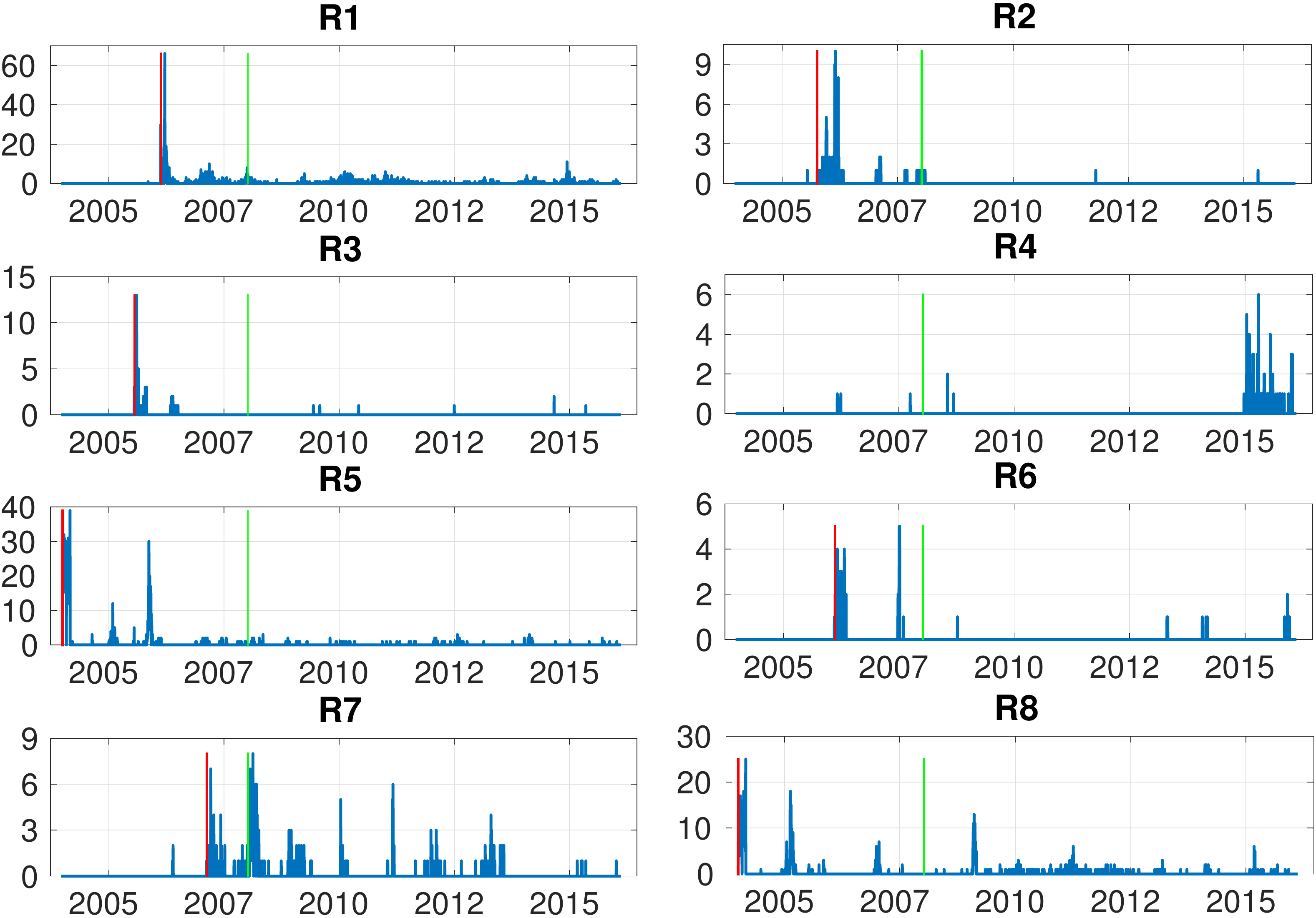}
\caption{Observed Time Series in the Impacted Regions}
\label{fig:fig4}
\end{figure}

\begin{figure}[!htb]
\centering
\subfloat[Configuration 1, Weight= $48 \%$]{\includegraphics[trim={0mm 1mm 1mm 1mm},clip,width= 0.22 \textwidth, height=0.22 \textwidth]{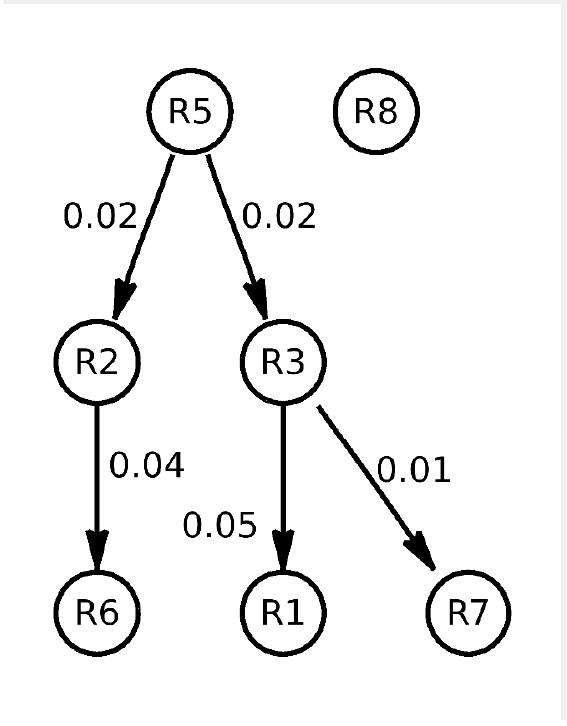}\label{fig:fig1_first_case}}\hspace*{0.05in}
\subfloat[Configuration 2, Weight= $23 \%$]{\includegraphics[trim={0mm 1mm 1mm 1mm},clip,width= 0.22 \textwidth, height=0.22 \textwidth]{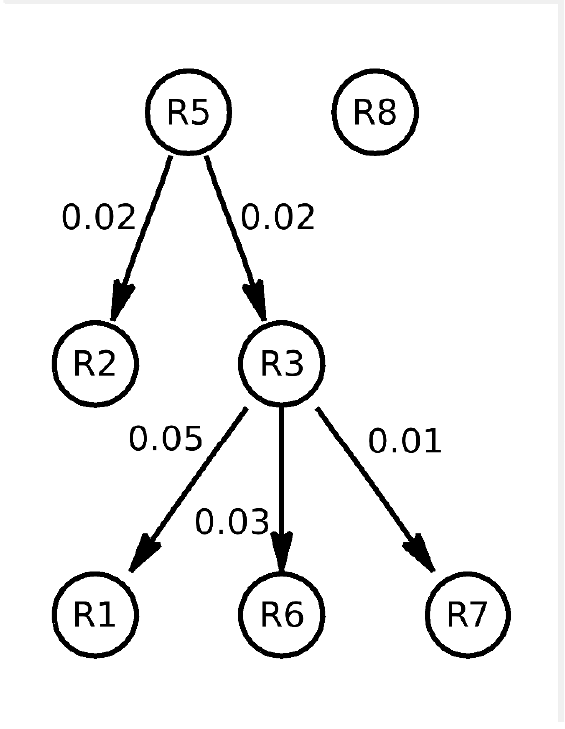}\label{fig:fig1_third_case}}\\
\subfloat[Configuration 3, Weight= $17 \%$]{\includegraphics[trim={0mm 1mm 1mm 1mm},clip,width= 0.22 \textwidth, height=0.22 \textwidth]{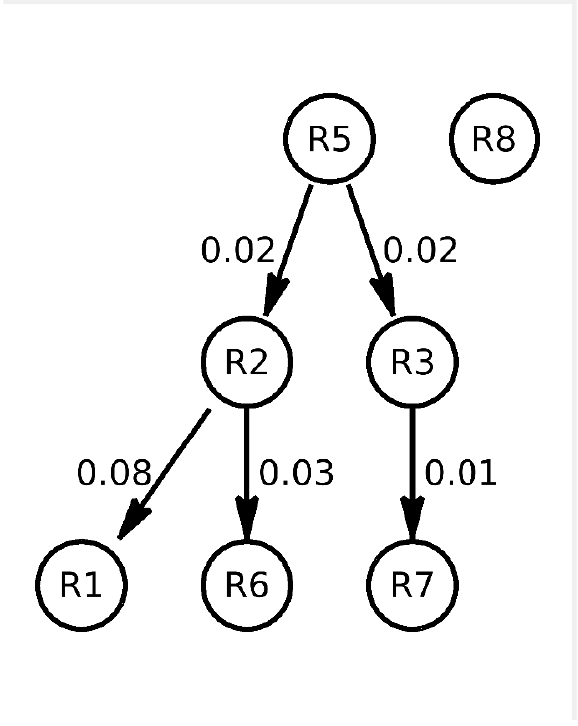}\label{fig:fig1_first_case}}\hspace*{0.05in}
\subfloat[Configuration 4, Weight= $10 \%$]{\includegraphics[trim={0mm 1mm 1mm 1mm},clip,width= 0.22 \textwidth, height=0.22 \textwidth]{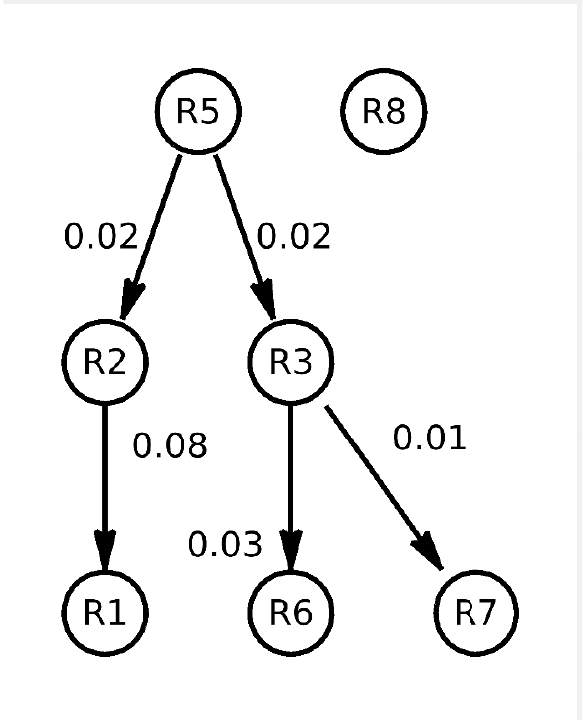}\label{fig:fig1_third_case}}
\caption{Most Possible Network Configurations}
\label{fig:fig5}
\end{figure}

We study the outbreak of Avian Influenza (H5N1 HPAI)
\cite{EMPRES}. Figure \ref{fig:fig3} shows the observed locations of
reported infections for both domestic and wild bird species for the
period of January 2004 to February 2016. We divide the observation
points to eight main regions using K-means clustering and generate a
time series $\bd_i$ to the $i$th region. The value of this time series at
day $n$ ($d_i^n$) denotes the number of separate locations within the region
$i$ in which the disease was reported on that day.
We model the number of observations in each region by a Poisson distribution:
\begin{equation}
f(\bd_i|t_i)=\prod_{n=1}^{t_i-1} \frac{{\lambda_{1i}}^{d_i^n} e^{-\lambda_{1i}}}{d_i^n!}\prod_{n=t_i}^T \frac{{\lambda_{2i}}^{d_i^n} e^{-\lambda_{2i}}}{d_i^n!}
\end{equation}
where $\lambda_{1i}=\frac{\sum_{n=t_1}^{t_i}d_i^n}{t_i-1}$ and
$\lambda_{2i}=\frac{\sum_{n=t_i+1}^T d_i^n}{T-t_i+1}$. The link
strength parameters $a_{ij}$ and $b_{ij}$ of equation
\eqref{prior_alpha} are derived by fitting a gamma distribution to the
inverse of distances between observation points of regions $i$ and
$j$. Figure \ref{fig:fig4} shows the time series for the eight
regions. Regions R5 and R8 are the first regions in which the disease
is observed. The first infections for these regions were reported on
the same day, so we assume that they were both sources of the infection.
We infer the infection parameters for the period 2004-2007
by generating $M=10^6$ samples and discarding the first $10^4$ ones. The green line in Figure
\ref{fig:fig4} shows the end of the study period. Region R4 has almost
no reported infections for this period so we exclude it when estimating the
underlying infection graph. The detected infection times are shown in Figure
\ref{fig:fig4} by red vertical lines.
Figure \ref{fig:fig5} shows the four most probable configurations of
the infection network and their percentages among generated samples.  The
edge weights in these graphs are estimated link strengths.
\section{CONCLUSION}
\label{sec:conc}

In this paper, we have proposed a framework for inferring the
underlying graph based on which an infection is diffused in a network
structure. We designed the model to address scenarios where the
infection times are unknown. We evaluated the performance using
synthetic datasets, demonstrating that (i) the incorporation of the
model could improve the estimation of infection times compared to
univariate changepoint estimation when the data match the model; and
(ii) the absence of exact knowledge of infection times does not lead
to significant deterioration in performance. We illustrated how the model and inference methodology could be applied to analyze the
outbreak of a virus. Incorporating multiple changepoint detection approaches can be studied as a future work. %In future work we aim to incorporate multiple changepoint detection approaches in order to account for infection recoveries and multiple changes of node state.

\bibliographystyle{IEEEtran}
\bibliography{refs}

\end{document}